\documentclass[conference]{IEEEtran}
\IEEEoverridecommandlockouts

\usepackage{graphicx}
\usepackage{enumerate}
\usepackage{cite}
\usepackage{bm}
\usepackage{color}
\usepackage{amsmath,amssymb,amsthm}
\usepackage[linesnumbered,boxruled,vlined]{algorithm2e}
\usepackage[hidelinks]{hyperref}

\usepackage{config}

\usepackage[font=footnotesize,skip=2pt]{caption}
\begin{document}

\title{Malicious RIS Meets RSMA: Unveiling the Robustness of Rate Splitting to RIS-Induced Attacks}

\author{\IEEEauthorblockN{Arthur S. de Sena\IEEEauthorrefmark{2}, André Gomes\IEEEauthorrefmark{3}, Jacek Kibi\l{}da\IEEEauthorrefmark{3}, Nurul H. Mahmood\IEEEauthorrefmark{2},\\ Luiz A. DaSilva\IEEEauthorrefmark{3}, Matti Latva-aho\IEEEauthorrefmark{2} }
	\IEEEauthorblockA{\hspace{3mm}
	 \IEEEauthorrefmark{2} University of Oulu, 6G Flagship, Finland \hspace{6mm} \IEEEauthorrefmark{3} Virginia Tech, CCI, USA\vspace{2mm}}
	{Emails: \{arthur.sena, nurulhuda.mahmood, matti.latva-aho\}@oulu.fi, \{gomesa, jkibilda, ldasilva\}@vt.edu}
\vspace{0mm}
\thanks{André is now with Rowan University, USA. Email: gomesa@rowan.edu.}
}

\maketitle
\begin{abstract}
    While the robustness of \ac{RSMA} to imperfect \ac{CSI} is well-documented, its susceptibility to attacks launched with malicious \acp{RIS} remains unexplored. This paper fills this gap by investigating three potential \ac{RIS}-induced attacks against \ac{RSMA} in a multi-user \ac{MIMO} network: random interference, aligned interference, and mitigation attack. The random interference attack employs random \ac{RIS} coefficients to disrupt \ac{RSMA}. The other two attacks are triggered by optimizing the \ac{RIS} through weighted-sum strategies based on the projected gradient method. Simulation results reveal significant degradation caused by all the attacks under perfect \ac{CSI} conditions. Remarkably, when imperfect \ac{CSI} is considered, \ac{RSMA}, owing to its flexible power allocation strategy designed to counter \ac{CSI}-related interference, can be robust to the attacks even when the base station is blind to them. It is also shown that \ac{RSMA} can significantly outperform conventional \ac{SDMA}.
\end{abstract}



\acresetall

\section{Introduction}



Next-generation \ac{MIMO} systems, employing an ever-increasing number of antennas, are a central pillar for future \ac{6G} wireless networks, as they allow for simultaneous transmissions to multiple users under the same frequency and time slot. One of the major challenges lies in maintaining low inter-user interference while serving multiple users. Conventional approaches employ linear precoding strategies to implement \ac{SDMA}, which requires accurate \ac{CSI}. In practical systems, only imperfect \ac{CSI} is available, which inevitably leads to residual inter-user interference at the receiver~\cite{Clerckx23}.

To overcome the limitations of \ac{SDMA}, \ac{RSMA} was proposed~\cite{Clerckx23, Sena22, Dai16, Joudeh16, Mao18}. \ac{RSMA} implements a flexible split transmission scheme aided by \ac{SIC} at the receiver side. At the \ac{BS}, one part of the users' data is encoded into private messages and transmitted via private precoders, while the other part is encoded into a common message and broadcast through a common precoder \cite{Clerckx23, Sena22}. The private precoders are designed similarly as in \ac{SDMA}, making them sensitive to imperfect \ac{CSI}. In contrast, the common precoder is constructed as a multicast precoder, which increases the system's robustness to inter-user interference.

The reliance of multi-user \ac{MIMO} systems on \ac{CSI} makes them susceptible to attacks that alter the propagation environment~\cite{huang2023disco, huang2023illegal,Sena24,lyu2020irs}. These attacks may employ a \ac{RIS}, a low-power planar array of nearly passive reconfigurable elements that can be dynamically controlled to adjust the propagation environment. While \ac{RIS} is primarily seen as a performance-enhancing technology~\cite{Sena22}, it may also trigger powerful attacks against wireless links~\cite{huang2023disco, huang2023illegal,Sena24,lyu2020irs,staat2022mirror, chen2022malicious, wang2022wireless, Wang24}. In the \ac{SDMA} case, among other threats, \acp{RIS} can disrupt the channel estimation process and boost \ac{CSI} inaccuracy, making linear precoders incapable of tackling inter-user interference. Different malicious attack schemes have been identified, including attacks with random \ac{RIS} coefficients \cite{huang2023disco, huang2023illegal}, optimized attacks in which the \ac{RIS}-associated channels are aligned to boost interference~\cite{Sena24}, and cancellation attacks, where the \ac{RIS} is optimized to add up the direct and reflected signals destructively at the receiver \cite{lyu2020irs}. While the robustness of \ac{RSMA} to interference originating from \ac{CSI} imperfections has been demonstrated~\cite{Clerckx23, Dai16, Mao18}, its susceptibility to \ac{RIS}-induced attacks remains an open question.

This paper delves for the first time into potential adversarial attacks that can be launched with the help of a malicious \ac{RIS} against \ac{RSMA}. We investigate a downlink multi-user \ac{MIMO} network, in which a nearby attacker controls an adversarial \ac{RIS}, as can be seen in \Fig{f1}. For this scenario, the attacker exploits the training protocol employed at the \ac{BS} to execute three different \ac{RIS} attack options not yet explored in the context of \ac{RSMA}: \emph{random interference}, \emph{aligned interference}, and \emph{mitigation attack}. The \emph{random interference attack} attempts to degrade the transmission of common and private data messages by configuring the \ac{RIS} with random phase shifts. The \emph{aligned interference attack} tries to maximize the reflected power by exploiting the channels associated with the \ac{RIS} to further diminish the effectiveness of \ac{RSMA} precoders. Finally, the \emph{mitigation attack} attempts to minimize the signal power at the users by destructively adding the reflected RIS channels to the legitimate direct user channels. For the latter two attacks, we consider weighted-sum strategies based on the projected gradient method to optimize the adversarial \ac{RIS} phase shifts.  

Our numerical results reveal a remarkable property of \ac{RSMA} that manifests itself in scenarios with imperfect \ac{CSI}. By flexibly allocating power to the common message, \ac{RSMA} can (unintentionally) mitigate the impact of the attacks, considerably outperforming \ac{SDMA} in all the considered threat scenarios under imperfect CSI. Counterintuitively, the proposed attacks have the strongest impact under perfect \ac{CSI} conditions. The results show that the attacks can potentially cause significant impact in all scenarios, with the severest performance degradation observed for the mitigation attack.

\vspace{2mm}
\noindent  {\it Notation:}
 Boldface lower-case letters denote vectors and upper-case represent matrices. The $i$th column of a matrix $\mathbf{A}$ is denoted by $[\mathbf{A}]_{:,i}$, the transpose and Hermitian transpose of $\mathbf{A}$ are represented by $\mathbf{A}^T$ and $\mathbf{A}^H$, respectively, $\mathbf{1}_{M, N}$ is the $M\times N$ all-ones matrix, and $\diamond$ represents the Khatri-Rao product. The operator $\mathrm{vec}\{\cdot \}$ transforms an $M\times N$ matrix into a column vector, $\mathrm{vecd}\{\cdot \}$ converts the diagonal elements of an $M\times M$ matrix into a column vector, $\mathrm{diag}\{\cdot \}$ transforms a vector of length $M$ into an $M\times M$ diagonal matrix, and $\angle(z)$ returns the phase of the complex scalar $z$.

\section{System Model}
\begin{figure}[t]
	\centering
	\includegraphics[width=.9\linewidth]{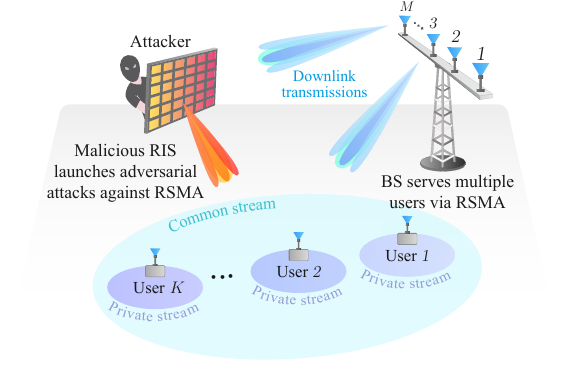}
	\caption{An attacker deploys an \ac{RIS} to perform adversarial attacks against RSMA in a downlink multi-user \ac{MIMO} system.}\label{fig:f1}
\end{figure}

This paper studies the downlink \ac{MIMO} system illustrated in \Fig{f1}, where one \ac{BS} employing $M$ antennas performs downlink data transmission to $K$ single-antenna users, represented as $\mathcal{K} = \{1,2,\cdots, K\}$, utilizing \ac{RSMA}. We assume an attacker deploys one malicious \ac{RIS} with $L$ reflecting elements. For the attacks to be effective, the \ac{RIS} is set to an absorption mode during the channel estimation phase and turned on only when the data transmission starts, as considered in \cite{huang2023disco, huang2023illegal, Sena24}. 

We adopt in this work the single-layer \ac{RSMA}, where the \ac{BS} transmits a single common message and users are required to carry out a single-layer  \ac{SIC} \cite{Clerckx23}. Under the single-layer \ac{RSMA} protocol, the message for each user is first split into common and private parts at the \ac{BS}. All the users' common parts are then encoded into a single common super symbol $x^c$, while the private parts are individually mapped into private symbols $x^p_{k}$. Next, the common and private symbols $x^c$ and $x^p_{k}$ are linearly precoded and superimposed in the power domain for transmission, resulting in the following data vector
\begin{align}\label{trans_sig}
    \mathbf{x} &= \mathbf{p}^c  \sqrt{P \alpha^c} x^c + \sum_{k=1}^{K} \mathbf{p}^{p}_{k} \sqrt{P \alpha_{k}^p} x_{k}^p  \in \mathbb{C}^{M},
\end{align}
where $P$ is the total transmit power, $\alpha^c$ and $\alpha_{k}^p$ are the power allocation coefficients for the common and private symbols, and $\mathbf{p}^{c} \in \mathbb{C}^{M}$ and $\mathbf{p}^{p}_{k} \in \mathbb{C}^{M}$ are the linear precoding vectors responsible for transmitting the corresponding symbols. 

\subsection{Signal reception and performance metrics}

After transmission, the superimposed RSMA data streams propagate through the direct link and the reflected one via the malicious \ac{RIS}. As a result, the $k$th user will receive
\begin{align}\label{rec_sig_01}
y_{k} = \big( \mathbf{f}_{k}^H
      \bm{\Theta}
    \mathbf{G} +  
    \mathbf{h}_{k}^H\big) \hspace{-1mm}\left( \mathbf{p}^c  \sqrt{P \alpha^c} x^c 
    + \hspace{-1mm} \sum_{i=1}^{K} \mathbf{p}^{p}_{i} \sqrt{P \alpha_{i}^p} x_{i}^p\right)\hspace{-1mm} + n_{k}, 
\end{align}
where $\mathbf{h}_{k} \in \mathbb{C}^{M}$, $\mathbf{G} \in \mathbb{C}^{L \times M}$, and $\mathbf{f}_{k} \in \mathbb{C}^{L}$ model the wireless channels between the BS and the $k$th user (link BS-U), the BS and the RIS (link BS-RIS), and the RIS and the $k$th user (link RIS-U), respectively. $\bm{\Theta} = \mathrm{diag}\{\mu_1 e^{-j \theta_1}, \cdots, \mu_L e^{-j \theta_L}\} \in \mathbb{C}^{L\times L}$ is the diagonal matrix modeling the reflections induced by the malicious \ac{RIS}, satisfying $|\mu_l|^2 = 1$ and $\theta_l \in [0, 2\pi], \forall l = 1, \cdots, L$, and $n_{k} \in \mathbb{C}$ is the noise coefficient for the $k$th user, following the complex Gaussian distribution with zero mean and variance $\sigma^2$.

The common message is recovered directly from \eqref{rec_sig_01} while the private messages are treated as noise. As a result, the common message will be decoded by the $k$th user with the following \ac{SINR}
\begin{align}\label{common_sinr}
    \gamma^c_k = \frac{|\big(\mathbf{f}_{k}^H\bm{\Theta}\mathbf{G} +  \mathbf{h}_{k}^H\big) \mathbf{p}^c|^2 P \alpha^c}{\sum_{i=1}^{K} | \big(\mathbf{f}_{k}^H\bm{\Theta}\mathbf{G} +  \mathbf{h}_{k}^H\big) \mathbf{p}^{p}_{i}|^2 P \alpha_{i}^p + \sigma^2},
\end{align}
where the term in the denominator accounts for the interference generated by both the $k$th intended private message and unintended private messages, resulting from imperfect \ac{CSI} used in the precoder design and the malicious \ac{RIS} attack. The instantaneous common rate experienced at the $k$th user is then given by $R_k^c = \log_2 (1 + \gamma_k^c)$. Since all users need to decode the common message, the actual allocated rate will be $R^c = \min_{\forall k \in \mathcal{K}}\{R_k^c\}$.

Once the common message is retrieved, \ac{SIC} is executed to subtract it from \eqref{rec_sig_01}. We assume that \ac{SIC} can successfully remove the interference associated with the common message. However, the $k$th user still experiences residual interference of private messages intended for other users $k' \neq k \in \mathcal{K}$, i.e., due to \ac{CSI} error and \ac{RIS} interference. Thus, the private \ac{SINR} for the $k$th user will be
\begin{align}\label{private_sinr}
    \gamma^p_k = \frac{|\big(\mathbf{f}_{k}^H\bm{\Theta}\mathbf{G} +  \mathbf{h}_{k}^H\big) \mathbf{p}^{p}_{k}|^2 P \alpha_{k}^p}{\sum_{i=1, i\neq k}^{K} | \big(\mathbf{f}_{k}^H\bm{\Theta}\mathbf{G} +  \mathbf{h}_{k}^H\big) \mathbf{p}^{p}_{i}|^2 P \alpha_{i}^p + \sigma^2},
\end{align}
resulting in the following achievable private rate $R_k^p = \log_2 (1 + \gamma_k^p)$. Consequently, the sum rate, in bits/s/Hz, experienced in the system can be expressed as
   $ R = R^c + \sum_{k=1}^K R_k^p $.


\subsection{CSI acquisition and precoder design at the BS}
During the channel estimation phase, the attacker sets its \ac{RIS} to an absorption mode so that the \ac{BS} does not account for the reflected \ac{RIS} channels in the estimation process. Nevertheless, we assume that the estimate of the legitimate fast-fading channels, $\mathbf{h}_{k}, \forall k \in \mathcal{K}$, acquired at the \ac{BS} is imperfect. This implies that the \ac{BS} precoders are designed based on a corrupted version of $\mathbf{h}_{k}$, modeled by \cite{Dai16}
\begin{align}\label{eq_imp_csi}
    \hat{\mathbf{h}}_{k} = \sqrt{1 - (\tau^{\text{\tiny BS-U}})^2} \mathbf{h}_{k} + \tau^{\text{\tiny BS-U}} \mathbf{z}_{k},
\end{align}
where $\mathbf{z}_{k}$ is the error vector independent of $\mathbf{h}_{k}$, whose entries follow the complex Gaussian distribution with zero mean and unit variance, and the coefficient $\tau^{\text{\tiny BS-U}} \in [0,1]$ models the quality of the \ac{CSI} estimation.

Given the above considerations, the private precoding vector $\mathbf{p}^{p}_{k} \in \mathbb{C}^{M}$ should be designed to ensure the (ideally) interference-free delivery of the private messages. More specifically, we wish to achieve $\big[\mathbf{h}_{k'}^H\big] \mathbf{p}^{p}_{k} \approx 0$, $ \forall k' \neq k \in \mathcal{K}$. Such a goal can be accomplished by designing $\mathbf{p}^{p}_{k}$ as a zero-forcing precoder based on the acquired estimate of $\mathbf{h}_{k}$ in \eqref{eq_imp_csi}, as follows. First, let us define $\hat{\mathbf{H}} = \big[\hat{\mathbf{h}}_{1}, \hat{\mathbf{h}}_{2}, \cdots, \hat{\mathbf{h}}_{K}\big] \in \mathbb{C}^{M \times K}$. Then, the private precoder for the $k$th user can be given by
\begin{align}\label{private_prec}
\mathbf{p}^{p}_{k} = \big[ \mathbf{\hat{H}} \big( \mathbf{\hat{H}}^H \mathbf{\hat{H}} \big)^{-1} \big]_{:,k} \in \mathbb{C}^{M},    
\end{align}
where it must be ensured that $M \geq K$.

As for the common precoder, $\mathbf{p}^{c} \in \mathbb{C}^{M}$, it should offer a satisfactory reception of the common message to all users.
However, 
the design of such precoders generally leads to NP-hard problems that can only be solved sub-optimally through, conventionally, iterative methods with high computational complexity
, e.g., semidefinite relaxation and successive convex approximation-based methods 
\cite{Konar17}. Alternatively, as in \cite{Dai16}, we adopted a lower complexity weighted \ac{MF} strategy, which can be computed in closed form by
\begin{align}\label{common_prec}
    \mathbf{p}^{c} = \sum_{i=1}^{K} \mu_k \hat{\mathbf{h}}_{i},
\end{align}
where $\mu_k$ is the weight for the $k$th user. In particular, the weights are adjusted as $\mu_1 = \cdots = \mu_K = \frac{1}{\sqrt{\mathbf{s}^H \mathbf{s}} }$, with $\mathbf{s} = \sum_{i=1}^{K} \hat{\mathbf{h}}_{i}$, such that $\|\mathbf{p}^{c}\|_2^2 = 1$.
The precoding method in \eqref{common_prec} becomes asymptotically optimal as the number of transmit antennas grows large \cite{Dai16}.

\subsection{Power allocation}\label{pa_subsection}
This paper adopts a simple yet effective adaptive power allocation strategy inspired by \cite{Clerckx23}, where the power is allocated in such a way that the interference resulting from imperfect \ac{CSI} in the \ac{SINR} for the private messages in \eqref{private_sinr}, corresponding to the legitimate BS-U channels, reaches approximately the same level as the noise power, i.e., $\sum_{i=1, i\neq k}^{K} | \mathbf{h}_{k}^H \mathbf{p}^{p}_{i}|^2 P \alpha_{i}^p \approx \sigma^2$. Specifically, the power allocation coefficient for the common message is determined as a function of the power allocated to the private messages as $\alpha^{c} = 1 - \sum_{i=1}^K \alpha^{p}_{i}$, in which a uniform power allocation is employed across the private coefficients, such that $\alpha^{p} = \alpha^{p}_{1} = \cdots = \alpha^{p}_{K}$, based on the criteria:
\begin{align}\label{pa_2}
    \alpha^{p}  = \min \Bigg\{\frac{1}{K}, \frac{\sigma^2}{\underset{{\forall k}}{\min} \sum_{i=1, i\neq k}^{K} | \mathbf{h}_{k}^H \mathbf{p}^{p}_{i}|^2 P} \Bigg\}.
\end{align}

As can be noticed, the greater the interference levels generated by imperfect \ac{CSI}, the more power is allocated to the common message. As explained in \cite{Clerckx23}, making the interference power similar to the noise power ensures that the experienced data rates do not saturate in the high transmit power regime, i.e., $R \rightarrow \infty$ as $P \rightarrow \infty$ even under imperfect \ac{CSI} scenarios. Note that for the above approach to work, the noise and interference powers (or their ratio) must be reported by each user to the \ac{BS} in the training phase. More advanced allocation methods, such as in \cite{Joudeh16, Mao18}, are left for future work.

\section{Potential RIS-Induced Attacks against RSMA}

The malicious goal of the attacker is to compute a reflection matrix $\bm{\Theta}$ that induces a performance degradation of the employed \ac{RSMA} scheme, i.e., degrade the system sum rate.
%
%
In the following subsections, we investigate three approaches to accomplish the goal and analyze the necessary \ac{CSI} knowledge for their implementation. 
%

\subsection{Random interference attack}
The most straightforward strategy that can cause performance degradation in \ac{RSMA} consists of configuring randomly the RIS reflecting coefficients to launch passive jamming attacks, similar to the attack against \ac{SDMA} proposed in~\cite{huang2023disco}. Random RIS interference attacks should reduce the effectiveness of the precoders in \eqref{private_prec} and \eqref{common_prec}, consequently leading to degradation in the rates of both common and private messages, without the need for any \ac{CSI} knowledge. The strongest impact should be observed against the private messages, given that the associated rates will become interference-limited due to the inability of the private precoders to cancel out the inter-user interference propagating through the malicious \ac{RIS}. 
It is noteworthy that, even though \ac{CSI} is not needed to optimize the \ac{RIS}, the attacker must know when the channel estimation and power allocation training phases happen. With this information, the attacker can configure the \ac{RIS} to absorb impinging signals and avoid being detected by the \ac{BS}.
%

\subsection{Aligned interference attack}\label{aligned_attack}
In our recent work~\cite{Sena24}, we demonstrated that if the attacker manages to acquire at least the illegitimate BS-RIS and RIS-U channels, it becomes possible to optimize the \ac{RIS} to cause a powerful performance degradation. In this subsection, 
 we show how such an optimized interference attack can be extended to \ac{RSMA}. 

For this attack to be effective in \ac{RSMA}, as in the previous subsection, the \ac{RIS} is set to absorption mode for both the channel estimation and power allocation. Moreover, we assume that the attacker has enough computational power to estimate the channels of the links BS-RIS and RIS-U, which are modeled as $\hat{\mathbf{G}} = \sqrt{1 - (\tau^{\text{\tiny BS-RIS}})^2} \mathbf{G} + \tau^{\text{\tiny BS-RIS}} \tilde{\mathbf{Z}}$ and $\hat{\mathbf{f}}_k = \sqrt{1 - (\tau^{\text{\tiny RIS-U}})^2} \mathbf{f}_{k} + \tau^{\text{\tiny RIS-U}} \tilde{\mathbf{z}}_{k}$, respectively, where $\tilde{\mathbf{Z}}$ and $\tilde{\mathbf{z}}_{k}$ are the associated error matrix and vector with entries following the complex standard Gaussian distribution, and $\tau^{\text{\tiny BS-RIS}}$ and $\tau^{\text{\tiny RIS-U}}$ are the error coefficients of the corresponding links, similarly as in \eqref{eq_imp_csi}. 
Observe that the interference propagating through the cascade \ac{RIS} channels, visible in the \ac{SINR} of the common message in \eqref{common_sinr}, also impacts the private messages. This implies that aligning $\hat{\mathbf{G}}$ with the channel vector $\hat{\mathbf{f}}_k$ of users $k \in \mathcal{K}$ should cause degradation to the rates of both common and private messages. With access only to \ac{RIS}-associated \ac{CSI}, the attacker can launch an attack against \ac{RSMA} through the following weighted-sum maximization
\begin{subequations}\label{prob_2}
\begin{align}
    & \underset{\bm{\Theta}}{\arg \max} \hspace{2mm}
     \sum_{k=1}^{K} \omega_{k} \|\hat{\mathbf{f}}_{k}^H\bm{\Theta}\hat{\mathbf{G}}\|_2^2,
    \label{prob_2a}\\[-1mm]
    &\hspace{4mm} \text{s.t.} \hspace{6mm} \bm{\Theta} = \mathrm{diag}\{\mu_1 e^{-j \theta_1}, \cdots, \mu_L e^{-j \theta_L}\},\label{prob_2b}\\[-1mm]
    & \hspace{14mm} |\mu_l|^2 = 1, \hspace{1mm} \forall l \in\{ 1,\cdots, L\}.\label{prob_2c}
\end{align}
\end{subequations}
where $\omega_k$ are weights that can be exploited to set the intensity of the attacks against each user. The current matrix structure of \eqref{prob_2} is challenging to tackle. To achieve a tractable version of the problem, the attacker invokes the following Khatri-Rao identity: $\left(\mathbf{Z}^T \diamond \mathbf{X} \right) \mathrm{vecd}\{\mathbf{Y}\} = \mathrm{vec}\{ \mathbf{X}\mathbf{Y}\mathbf{Z}\}$. By relying on this property, we can define
  $\bm{\theta} \triangleq \mathrm{vecd}\{\bm{\Theta}\} \in \mathbb{C}^{L}$ and
   $\hat{\mathbf{K}}_{k} \triangleq \hat{\mathbf{G}}^T \diamond \hat{\mathbf{f}}_{k}^H \in \mathbb{C}^{M \times L}$. 
These transformations are then applied to \eqref{prob_2}, resulting in the following simpler problem
\begin{subequations}\label{prob_3}
\begin{align}
    & \underset{\bm{\theta}}{\arg \max} \hspace{2mm} 
     \sum_{k=1}^{K} \omega_{k} \left\| \hat{\mathbf{K}}_{k}\bm{\theta}
    \right\|_2^2,
    \label{prob_3a}\\[-1mm]
    &\hspace{4mm} \text{s.t.}  \hspace{6mm} |\mu_{l}|^2 = 1, \hspace{1mm} \forall l \in\{ 1,\cdots, L\}.\label{prob_3b}
\end{align}
\end{subequations}
Lastly, the matrices in the objective function in \eqref{prob_3a} are stacked to obtain the following equivalent problem
\begin{subequations}\label{prob_4}
\begin{align}
    & \underset{\bm{\theta}}{\arg \max} \hspace{1.5mm} 
     \left\| \begin{bmatrix} \sqrt{\omega_{1}} \hat{\mathbf{K}}_{1}^H &
     \hdots &
     \sqrt{\omega_{K}} \hat{\mathbf{K}}_{K}^H
    \end{bmatrix}^H \bm{\theta} \right\|_2^2,
    \label{prob_4a}\\[-1mm]
    &\hspace{4mm} \text{s.t.}  \hspace{6mm} |\mu_{l}|^2 = 1, \hspace{1mm} \forall l \in\{ 1,\cdots, L\}.\label{prob_4b}
\end{align}
\end{subequations}
Despite the non-convex element-wise modulus constraint in \eqref{prob_4b}, the desired solution can be efficiently approximated through a projected gradient method, as in \cite{Tranter17, Sena24}, which is presented in Algorithm \ref{alg1}, in which $\lambda_{\text{max}}(\cdot)$ computes the largest eigenvalue of a given matrix, $\Delta$ is the step size in the direction of the gradient, $\delta$ is a coefficient that controls the step size, and $I$ denotes the number iterations.

\begin{figure}
\centering
\begingroup
\csname @twocolumnfalse\endcsname
\resizebox{.46\textwidth}{!}{%
\begin{minipage}{.56\textwidth}
\setlength{\algomargin}{5mm}
\setlength{\interspacetitleboxruled}{1mm}
\begin{algorithm}[H]
	\SetKwRepeat{Do}{do}{while}
	\SetAlgoLined
	\KwIn{\small $I, \delta \in (0,1), \{\omega_1, \cdots, \omega_K\}, \{\hat{\mathbf{K}}_1, \cdots, \hat{\mathbf{K}}_K\}$\;\vspace{1mm}}

	\small
	Initialize: $\bm{\theta}_{(1)} = \mathbf{1}_{L,1}$, $\Delta = \delta/\lambda_{\text{max}}\left(\mathbf{\bar{K}}^H \mathbf{\bar{K}}\right)$, $ 
  \mathbf{\bar{K}} = \begin{bmatrix} \sqrt{\omega_{1}} \hat{\mathbf{K}}_{1}^H &
     \hdots &
     \sqrt{\omega_{K}} \hat{\mathbf{K}}_{K}^H
    \end{bmatrix}^H$\;\vspace{1mm}
        
		\For{$i = 1, 2, \cdots, I - 1$}{\Indmm 
		
		Update in the direction of the gradient of \eqref{prob_4a}: \hspace{50mm}
        $\bm{\vartheta} = \bm{\theta}_{(i)} + \Delta \mathbf{\bar{K}}^H\mathbf{\bar{K}} \bm{\theta}_{(i)}$\;\vspace{1mm}
				
		Compute the projection onto the unit 1-sphere:\hspace{50mm}
        $\bm{\theta}_{(i+1)} = e^{j\angle\left( \bm{\vartheta} \right)}$\;
		
		}\vspace{2mm}\KwOut{\small$\bm{\Theta} = \mathrm{diag}\{\bm{\theta}_{(I)}\}$.}
  
	\caption{\hspace{-1mm} RIS-induced interference attack against RSMA}\label{alg1}
\end{algorithm}
\end{minipage}
}%
\endgroup \vspace{-2mm}
\end{figure}

\subsection{Mitigation attack}
By inspecting the expressions \eqref{common_sinr} and \eqref{private_sinr}, it can be noticed that in the extreme case with no power allocated to the private messages, no interference will impact the \ac{SINR} of the common message. In such a scenario, the worst effect that the attacks from the previous subsections can cause is to make the wireless channels mismatched with the precoder in \eqref{common_prec}, i.e., due to the unexpected contribution of the reflected RIS channels. Even though this might lead to performance degradation, the data rates on the common message are not interference-limited. 
In this case, the attacker must opt for an attack capable of mitigating the common signal to create a stronger impact. To this end, the \ac{RIS} needs to be optimized such that reflected channels add destructively with the direct BS-U channels. This can be accomplished with the following minimization problem
\begin{subequations}\label{prob_5}
\begin{align}
    & \underset{\bm{\Theta}}{\arg \min} \hspace{2mm}
     \sum_{k=1}^{K} \omega_{k} \|\hat{\mathbf{f}}_{k}^H\bm{\Theta}\hat{\mathbf{G}} + \hat{\mathbf{h}}_{k}^H\|_2^2,
    \label{prob_5a}\\[-1mm]
    &\hspace{4mm} \text{s.t.} \hspace{6mm} \bm{\Theta} = \mathrm{diag}\{\mu_1 e^{-j \theta_1}, \cdots, \mu_L e^{-j \theta_L}\},\label{prob_5b}\\[-1mm]
    & \hspace{14mm} |\mu_l|^2 = 1, \hspace{1mm} \forall l \in\{ 1,\cdots, L\}.\label{prob_5c}
\end{align}
\end{subequations}

By relying on the Khatri-Rao property introduced in the last subsection, problem \eqref{prob_5} can be reformulated as follows
\begin{subequations}\label{prob_6}
\begin{align}
    & \underset{\bm{\theta}}{\arg \min} \hspace{2mm}
     \sum_{k=1}^{K} \omega_{k} \|\hat{\mathbf{K}}_{k}\bm{\theta} + (\hat{\mathbf{h}}_{k}^H)^T\|_2^2,
    \label{prob_6a}\\[-1mm]
    &\hspace{4mm} \text{s.t.} \hspace{6mm} |\mu_l|^2 = 1, \hspace{1mm} \forall l \in\{ 1,\cdots, L\},\label{prob_6b}
\end{align}
\end{subequations}
where $\bm{\theta}$ is the vector of reflecting coefficients, and $\hat{\mathbf{K}}_{k}$ is the matrix from the Khatri-Rao product, as in subsection \ref{aligned_attack}.

For solving \eqref{prob_6}, the terms of the sum in its objective function are stacked vertically so that the following is achieved
\vspace{-3mm}
\begin{subequations}\label{prob_7}
\begin{align}
    & \underset{\bm{\theta}}{\arg \min} \hspace{2mm}
    \left\|
    \begin{bmatrix}
    \sqrt{\omega_{1}} \hat{\mathbf{K}}_{1}^H \\
    \vdots \\
    \sqrt{\omega_{K}} \hat{\mathbf{K}}_{K}^H
    \end{bmatrix}
    \bm{\theta} + 
    \begin{bmatrix}
    \sqrt{\omega_{1}} (\hat{\mathbf{h}}_{1}^H)^T \\
    \vdots \\
    \sqrt{\omega_{K}} (\hat{\mathbf{h}}_{K}^H)^T
    \end{bmatrix}\right\|_2^2,
    \label{prob_7a}\\[1mm]
    &\hspace{4mm} \text{s.t.} \hspace{6mm} |\mu_l|^2 = 1, \hspace{1mm} \forall l \in\{ 1,\cdots, L\}.\label{prob_7b}
\end{align}
\end{subequations}

\begin{figure}
\centering
\begingroup
\csname @twocolumnfalse\endcsname
\resizebox{.46\textwidth}{!}{%
\begin{minipage}{.56\textwidth}
\setlength{\algomargin}{5mm}
\setlength{\interspacetitleboxruled}{1mm}
\begin{algorithm}[H]
	\SetKwRepeat{Do}{do}{while}
	\SetAlgoLined
	\KwIn{\small $I, \delta \in (0,1), \{\omega_1, \cdots, \omega_K\},$ $\{\hat{\mathbf{K}}_1, \cdots, \hat{\mathbf{K}}_K\}, \{\hat{\mathbf{h}}_1, \cdots, \hat{\mathbf{h}}_K\}$\;\vspace{1mm}}

	\small
	Initialize: $ 
  \mathbf{\bar{K}} = \begin{bmatrix} \sqrt{\omega_{1}} \hat{\mathbf{K}}_{1} &
     \hdots &
     \sqrt{\omega_{K}} \hat{\mathbf{K}}_{K}
    \end{bmatrix}^H$, $ 
  \mathbf{\bar{h}} = \begin{bmatrix} \sqrt{\omega_{1}} \hat{\mathbf{h}}_{1}^H &
     \hdots &
     \sqrt{\omega_{K}} \hat{\mathbf{h}}_{K}^H
    \end{bmatrix}^T$,
    $\bm{\theta}_{(1)} = e^{j\angle\left( [(\mathbf{\bar{K}}^H \mathbf{\bar{K}})^{-1}\mathbf{\bar{K}}^H] \mathbf{\bar{h}} \right)}$, $\Delta = \delta/\lambda_{\text{max}}\left(\mathbf{\bar{K}}^H \mathbf{\bar{K}}\right)$\;\vspace{1mm}
        
		\For{$i = 1, 2, \cdots, I - 1$}{\Indmm 
		
		Update in the opposite direction of the gradient of \eqref{prob_7a}: \hspace{50mm}
        $\bm{\vartheta} = \bm{\theta}_{(i)} - \Delta \mathbf{\bar{K}}^H (\mathbf{\bar{K}} \bm{\theta}_{(i)} + \bar{\mathbf{h}})$\;\vspace{1mm}
				
		Compute the projection onto the unit 1-sphere:\hspace{50mm}
        $\bm{\theta}_{(i+1)} = e^{j\angle\left( \bm{\vartheta} \right)}$\;
		
		}\vspace{2mm}\KwOut{\small$\bm{\Theta} = \mathrm{diag}\{\bm{\theta}_{(I)}\}$.}
  
	\caption{\hspace{-1mm} RIS-induced mitigation attack against RSMA}\label{alg2}
\end{algorithm}
\end{minipage}
}%
\endgroup \vspace{-3mm}
\end{figure}


The above problem is non-convex due to the element-wise unity modulus constraint. Still, like \eqref{prob_4}, this class of problems can be solved sub-optimally by exploiting a projected gradient strategy. The implemented approach is provided in Algorithm \ref{alg2}. Note that since all signals transmitted by the \ac{BS} propagate through the same channels, this attack should impact the detection of both common and private messages. Therefore, users will experience performance degradation independently of the amount of power allocated to either of the messages.
Consequently, in contrast to the other considered attacks, the mitigation scheme does not require the attacker to know when the power allocation training is carried out. 
This is important since for this adversarial scheme to be effective, the attacker needs the knowledge of estimates of the channels of the BS-RIS link, $\hat{\mathbf{G}}$, RIS-U link, $\hat{\mathbf{f}}_{k}$, and the legitimate BS-U link, $\hat{\mathbf{h}}_{k}^H$, similar to the attack proposed in~\cite{lyu2020irs}. Depending on the attacker's capabilities, obtaining the latter estimates may be costly or impractical. Thus, in our numerical results, we will evaluate a scenario where the attacker can only access imperfect channel estimates.

\section{Numerical Results}

In this section, we investigate the severity of the proposed attacks against \ac{RSMA} and the impact of different system parameters. We compare the performance of \ac{RSMA} and the conventional \ac{SDMA}, i.e., when $\alpha^c = 0$, to assess their robustness against the presented threats.

We implement the communication scenario where $K = 3$ single-antenna users are connected to a \ac{BS} with $M = 10$ transmit antennas. The coordinates of users $1, 2$, and $3$ are fixed at $(30, 15)$~m, $(50, 15)$~m, and $(55, 10)$~m, respectively, and the \ac{BS} at $(0, 0)$~m. Moreover, the attacker's \ac{RIS} comprises $L = 200$ reflecting elements and is deployed at the coordinate $(40, 5)$~m. With this setup, the path-loss coefficients are calculated as $(d^{\text{\tiny BS-RIS}})^{-\eta}$, $(d^{\text{\tiny RIS-U}}_{k})^{-\eta}$, and $(d^{\text{\tiny BS-U}}_{k})^{-\eta}$, for $k\in \{1,2,3\}$, where $d^{\text{\tiny BS-RIS}}$, $d_k^{\text{\tiny RIS-U}}$, and $d_k^{\text{\tiny BS-U}}$ represent the distances of the links BS-RIS, RIS-U, and BS-U, respectively, with $\eta$ denoting the path-loss exponent, set to $2.5$. As for Algorithms \ref{alg1} and \ref{alg2}, the step size parameter is configured as $\delta = 0.99$, the number of iterations as $I = 3\times 10^3$, and the weights are set to $\omega_1 = \omega_2 = \omega_3 = 1/3$. Furthermore, in the \ac{SDMA} scheme, the total transmit power of the \ac{BS} is allocated uniformly among the users, i.e., the fraction $P/K$ is allocated to each user, and the noise variance is set to $\sigma^2 = -50$~dBm.

\begin{figure}[t]
	\centering
	\includegraphics[width=.87\linewidth]{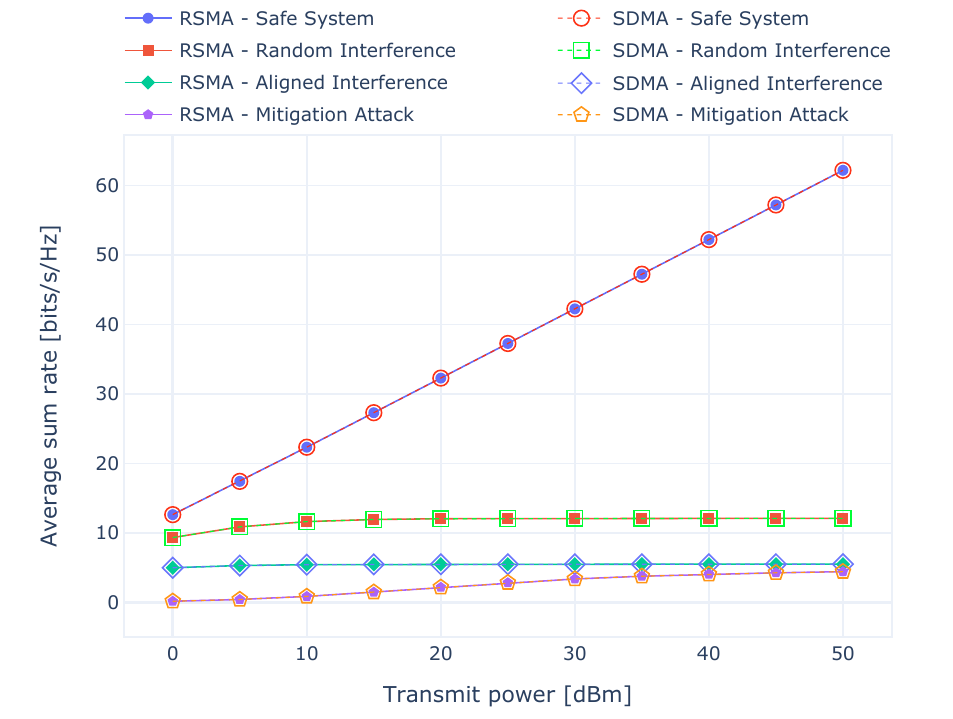}
	\caption{Average sum rate for different attack strategies with perfect CSI in all links, i.e., $\tau^{\text{\tiny BS-U}} = \tau^{\text{\tiny BS-RIS}} = \tau^{\text{\tiny RIS-U}} = 0$ at both the BS and the attacker. }\label{r1}\vspace{-3mm}
\end{figure}

We start with Fig. \ref{r1}, which presents sum rate curves for the case where both the BS and the attacker can estimate the channels perfectly. Because the BS has access to perfect CSI, the private precoders can completely remove inter-user interference. As a result, the power allocation strategy in subsection \ref{pa_subsection}, which cannot detect the RIS interference, will assign power primarily to the private messages, making RSMA perform identically as SDMA in all tested cases. As can be seen, in this ideal scenario with perfect CSI, all three kinds of RIS-induced attacks are able to severely deteriorate the performance of both RSMA and SDMA, with the random interference rendering the mildest impact, the aligned interference the second strongest, and the mitigation attack the strongest impact.

\begin{figure}[t]
	\centering
	\includegraphics[width=.87\linewidth]{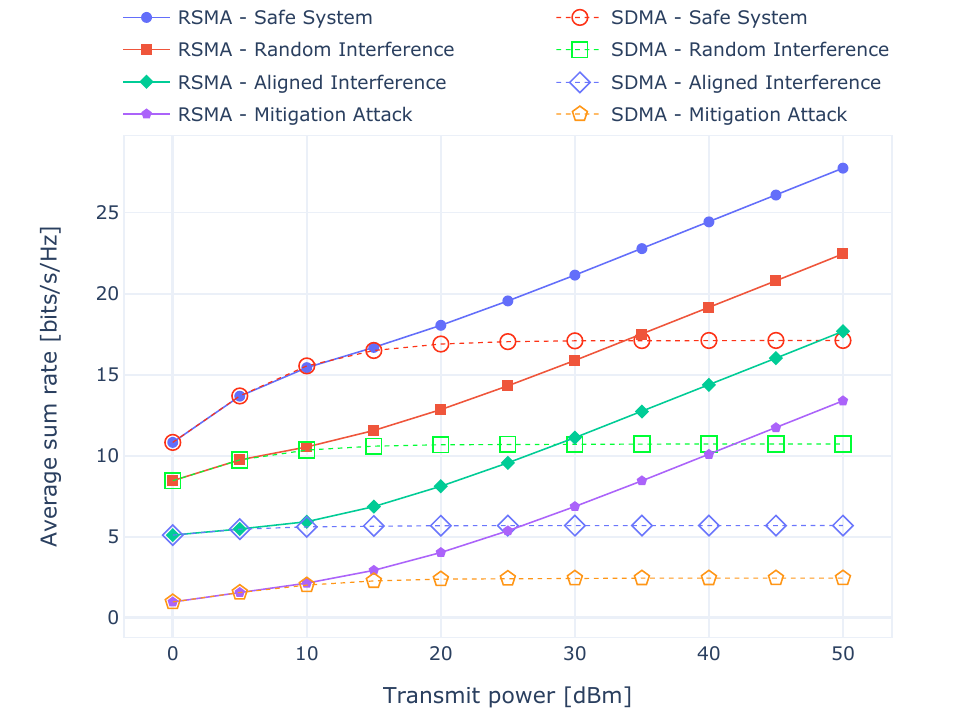}
	\caption{Average sum rate for different attack strategies with imperfect CSI in all links, with $ \tau^{\text{\tiny BS-U}} = \tau^{\text{\tiny BS-RIS}} = \tau^{\text{\tiny RIS-U}} = 0.3$ at both the BS and the attacker.}\label{r2}\vspace{-2mm}
\end{figure}

In Fig. \ref{r2}, we can visualize the behavior of the considered multiple access schemes for the scenario with imperfect CSI at both the BS and the attacker, considering an error factor of $ \tau^{\text{\tiny BS-U}} = \tau^{\text{\tiny BS-RIS}} = \tau^{\text{\tiny RIS-U}} = 0.3$. In this scenario, we see that the curves for the safe system are lower than those observed under perfect CSI in Fig. \ref{r1}. Moreover, the sum rate experienced with SDMA saturates as a consequence of the dominant inter-user interference in high transmit power values, either due to imperfect CSI or the RIS attacks, or both. On the other hand, the flexible power allocation policy of RSMA shifts power to the common message as the detected interference from imperfect CSI starts to grow. Note that, although the BS is blind to the RIS-induced interference, assigning power to the common message as a way to overcome degradation from imperfect CSI can also significantly alleviate (unintentionally) the impact of the attacks. Also, even though the different attacks can still cause performance degradation to RSMA, the sum rate curves are no longer interference-limited, i.e., the RSMA curves do not saturate as in the SDMA schemes. This confirms that RSMA is remarkably more robust than SDMA, even under such threat scenarios that may be difficult to detect. 

\begin{figure}[t]
	\centering
	\includegraphics[width=.9\linewidth]{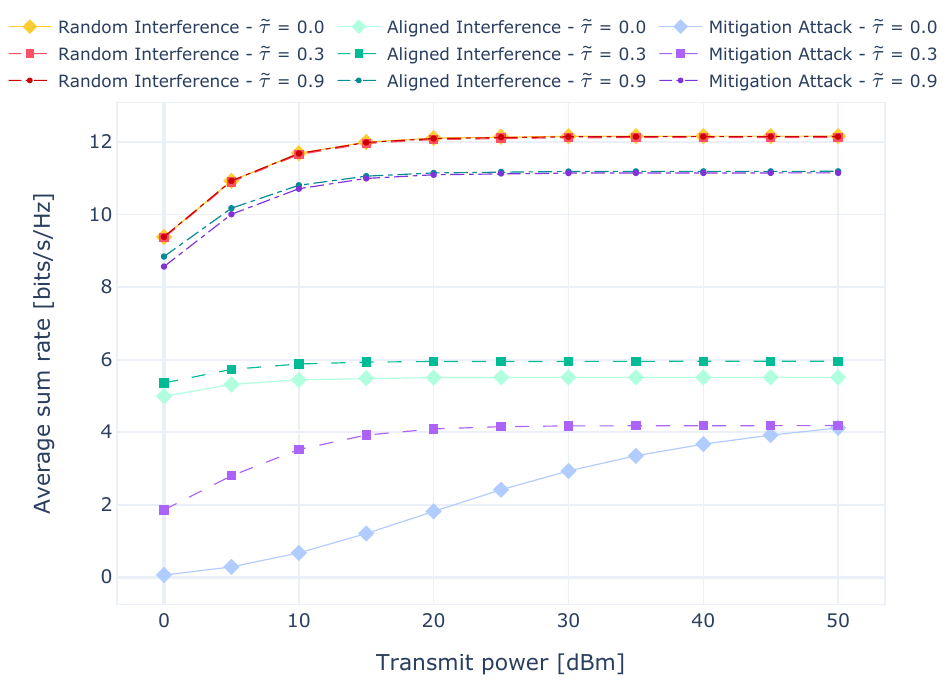}
	\caption{Average sum rate for RSMA under different attack strategies with perfect CSI at the BS, i.e., $\tau^{\text{\tiny BS-U}} = 0$ for the BS, and various CSI error levels at the attacker, considering $\Tilde{\tau} = \tau^{\text{\tiny BS-RIS}} = \tau^{\text{\tiny RIS-U}} = \tau^{\text{\tiny BS-U}}$ for the attacker.}\label{r3}\vspace{-3mm}
\end{figure}

\begin{figure}[t]
	\centering
	\includegraphics[width=.9 
 \linewidth]{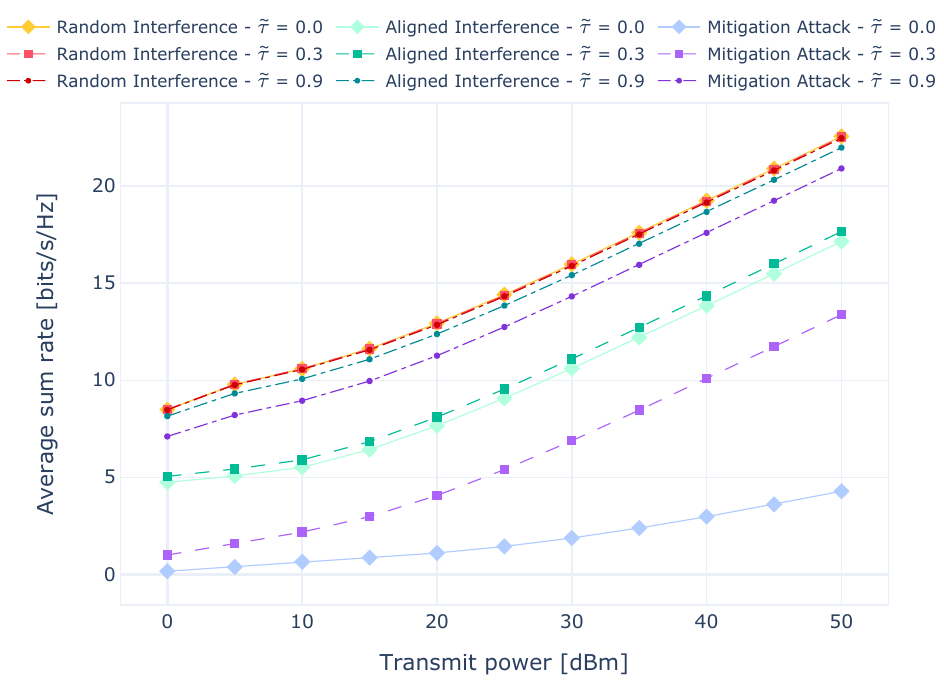}
	\caption{Average sum rate for RSMA under different attack strategies with $\tau^{\text{\tiny BS-U}} = 0.3$ for the BS, and various CSI error levels at the attacker, considering $\Tilde{\tau} = \tau^{\text{\tiny BS-RIS}} = \tau^{\text{\tiny RIS-U}} = \tau^{\text{\tiny BS-U}}$ for the attacker.}\label{r4}\vspace{-2mm}
\end{figure}

Lastly, Figs. \ref{r3} and \ref{r4} investigate the impact of the CSI quality on the attacks' severity. In Fig. \ref{r3}, specifically, we test different channel error values at the attacker, such that $\Tilde{\tau} = \tau^{\text{\tiny BS-RIS}} = \tau^{\text{\tiny RIS-U}} = \tau^{\text{\tiny BS-U}}$, while considering perfect CSI at the BS. The figure shows that the impact of both the aligned interference and the mitigation attack diminishes with the increase of the channel error factor, approaching the performance observed under the random interference attack as $\Tilde{\tau}$ gets large. Moreover, it is noteworthy that even with a high channel error, the mitigation attack remains the most impactful one, still slightly outperforming its aligned interference counterpart when $\Tilde{\tau} = 0.9$. It can also be observed
that, because the BS operates under perfect CSI, independently of the channel error at the attacker all attacks make the sum rate curves saturate in the high-power regime. This interference-limited behavior changes in Fig. \ref{r4}, which investigates different error levels at the attacker but considers imperfect CSI at the BS, with $\tau^{\text{\tiny BS-U}} = 0.3$. Again, the dominance of the mitigation attack persists even under high levels of error. These results indicate that if the attacker acquires at least imperfect estimates of both BS-U and RIS-induced channels, the mitigation strategy is the attack that poses the highest risk of performance degradation. Nevertheless, the flexible interference management of \ac{RSMA} can reverse the impacts of the attacks to some extent, as long as $ \tau^{\text{\tiny BS-U}} > 0$ at the BS. Counterintuitively, operating under imperfect \ac{CSI}, instead of being detrimental, makes \ac{RSMA} more robust to such adversarial \ac{RIS} attacks.

\section{Conclusions}
This paper covered three potential \ac{RIS}-induced attacks that can harm the performance of \ac{RSMA}, namely random interference, aligned interference, and mitigation attack. For the two latter attacks, we presented algorithms based on projected gradient methods that can efficiently find \ac{RIS} coefficients that lead to a strong degradation of the data rates of all connected users. Comprehensive simulation results demonstrated the severity of the different malicious schemes and revealed that \ac{RSMA} can be robust even when the \ac{BS} is blind to the attacks. We demonstrated that by smartly allocating power to the common message to avoid interference from imperfect \ac{CSI}, \ac{RSMA} can deliver data rates that considerably outperform \ac{SDMA} under the presented security threats. In future work, we will explore strategies for further improving the robustness of \ac{RSMA} and methods for countering such attacks.

\section*{Acknowledgments}
This work was supported by the Smart Networks and Services Joint Undertaking (SNS JU) under the European Union’s Horizon Europe research and innovation programme within \href{https://hexa-x-ii.eu/}{Hexa-X-II project} (grant no. 101095759), by Business Finland via the
6GBridge - Local 6G project (grant no. 8002/31/2022), and by 
the Research Council of Finland through the 6G Flagship (grant no. 346208) and the 6G-ConCoRSe project (grant no. 24300065). 
This work also received support from the \href{https://cyberinitiative.org/}{Commonwealth Cyber Initiative (CCI)} in Virginia, US, an investment in the advancement of cyber R\&D, innovation, and workforce development, and
by the National Science Foundation under grants no. 2318798 and 2326599.




\ifCLASSOPTIONcaptionsoff
\newpage
\fi

\bibliographystyle{IEEEtran}
\bibliography{main}

\end{document}